 \documentclass[12pt, draftclsnofoot, onecolumn]{IEEEtran}
\usepackage[latin9]{inputenc}
\usepackage{authblk}
\usepackage{array}
\usepackage{float}
\usepackage{mathtools}
\usepackage{amsmath}
\usepackage{amsthm}
\usepackage{amssymb}
\usepackage{graphicx}
\usepackage{wasysym}
\usepackage{setspace}
\usepackage{color}
\usepackage{bm}
\usepackage{cases}
\usepackage{booktabs}
\usepackage{tikz}
\usepackage{dblfloatfix}
\usepackage{flowchart}
\usetikzlibrary{matrix,shapes,arrows,positioning,chains}
\def\BibTeX{{\rm B\kern-.05em{\sc i\kern-.025em b}\kern-.08em
    T\kern-.1667em\lower.7ex\hbox{E}\kern-.125emX}}
\makeatletter

\floatstyle{ruled}
\newfloat{algorithm}{tbp}{loa}
\providecommand{\algorithmname}{Algorithm}
\floatname{algorithm}{\protect\algorithmname}

\theoremstyle{plain}

\theoremstyle{remark}

\theoremstyle{plain}

\theoremstyle{plain}

\usepackage{epsfig}
\usepackage{subfigure}
\usepackage{cite}
\usepackage{graphicx}
\usepackage{multirow}
\usepackage{array}
\usepackage{enumerate}
\usepackage{graphicx}
\usepackage{times}
\usepackage{algorithm}
\usepackage{tabularx}
\usepackage[noend]{algpseudocode}

\makeatletter
\let\OldStatex\Statex
\renewcommand{\Statex}[1][1]{%
  \setlength\@tempdima{\algorithmicindent}%
  \OldStatex\hskip\dimexpr#1\@tempdima\relax}
\makeatother

\algnewcommand{\Initialization}[1]{%
  \State \textbf{Initialization:} \hspace{0.13in}
}
\makeatletter
\newcommand{\multiline}[1]{%
  \begin{tabularx}{\dimexpr\linewidth-\ALG@thistlm}[t]{@{}X@{}}
    #1
  \end{tabularx}
}
\makeatother

\makeatother

\providecommand{\lemmaname}{Lemma}
\providecommand{\propositionname}{Proposition}
\providecommand{\remarkname}{Remark}
\providecommand{\theoremname}{Theorem}
\providecommand{\theoremname}{Definition}
\begin{document}

\title{Analysis of Channel Model for LAA-WLAN Coexistence with different OFDM Parameters between LAA and Wi-Fi}
\author{Harim Lee and Hyun Jong Yang,~\IEEEmembership{Member,~IEEE}
\thanks{Harim Lee and Hyun Jong Yang are with the School of Electrical and Computer Engineering, Ulsan National Institute of Science and Technology (UNIST), Ulsan, 44919, Korea (e-mail: \{hrlee,hjyang\}@unist.ac.kr).
Hyun Jong Yang is the corresponding author.}
}
\maketitle
\begin{abstract}
In this work, the channel model for the asynchronous Wi-Fi and LAA signals is investigated, also taking into consideration the impairment of the OFDM parameters between LAA and Wi-Fi.
Even for the same bandwidth, e.g., 20 MHz channel, OFDM parameters of LAA-LTE and Wi-Fi are different \cite{3GPPTS36_213_14,80211}. In addition, the legacy Wi-Fi devices cannot receive or transmit synchronously with LAA-LTE. 
Thus, the orthogonality between the subcarriers is broken if LAA-LTE's signals are received at Wi-Fi devices and vice versa. In order to facilitate an analysis of the throughput, we should derive the distortion of LAA-LTE's signals received at Wi-Fi devices and the distortion of Wi-Fi's signals received at LAA-LTE devices due to the difference in OFDM parameters of LAA-LTE and Wi-Fi. 
\end{abstract}

\section{Effective Channel Model for LAA-WLAN Coexistence}
\label{sec:Channel_Model}
According to each standard of LAA-LTE and Wi-Fi, both technologies adopt the orthogonal frequency-division multiplexing (OFDM) for each physical-layer but with different numbers of subcarriers, different sampling periods, and different cyclic prefix (CP) and data duration  even in the same bandwidth.
Hence, the orthogonality between the subcarriers is broken when LAA-LTE's signals are sampled at Wi-Fi devices and also vice versa. Moreover, since the transmission of LAA-LTE and Wi-Fi is initiated in an ad-hoc manner, frame and symbol synch is not guaranteed between LAA-LTE and Wi-Fi. As a result, we cannot derive per-subcarrier throughput with the previous analysis framework. 
In this section, we derive an impact of the difference in the physical-layer parameters on the effective channel. 

\begin{figure}[t]
\centering
\hspace{-0.1in}
\includegraphics[width=1\columnwidth]{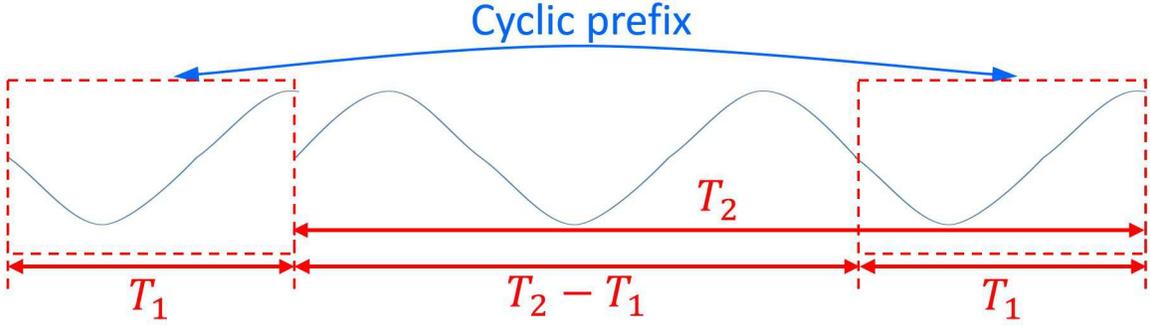}
\vspace{-0.12in}
\caption{Definition of the function $b(t; f,T_1, T_2)$ with the cyclic prefix.}
\label{Fig:b_function}
\end{figure}
\begin{table}
\resizebox{1\linewidth}{!}{
\begin{minipage}{\linewidth}
\caption{OFDM parameters of LAA-LTE and Wi-Fi for a 20 MHz channel}
\label{Table:OFDM_Paramters}
\centering
\begin{tabular}{m{0.55cm} m{0.6cm} m{6.4cm}}
\toprule
\multicolumn{1}{c}{LAA-}    & \multicolumn{1}{c}{\multirow{2}{*}{Wi-Fi}}      & \multicolumn{1}{c}{Description}\\
\multicolumn{1}{c}{LTE}    & \multicolumn{1}{c}{}      & \multicolumn{1}{c}{(Values for LAA-LTE and Wi-Fi)}\\ \hline
$f_{i}^\text{L}$               & $f_{i}^\text{W}$               & $i$-th subcarrier frequency\\
$\vartriangle \!\!\! f_\text{L}$ & $\vartriangle \!\! \!  f_\text{W}$ & Subcarrier spacing (15 kHz and 312.5 kHz)\\
$T_\text{data}^\text{L}$       & $T_\text{data}^\text{W}$       &  OFDM symbol duration (66.7$\mu$s and 3.2 $\mu$s)\\
$T_\text{CP}^\text{L}$         & $T_\text{CP}^\text{W}$         & Cyclic prefix duration (4.7 $\mu$s and 0.8 $\mu$s)\\
$T_\text{total}^\text{L}$      & $T_\text{total}^\text{W}$      & Total symbol duration (71.4 $\mu$s and 4 $\mu$s)\\
$N_\text{FFT}^\text{L}$        & $N_\text{FFT}^\text{W}$        & Number of subcarriers (2048 and 64)\\
$N_\text{CP}^\text{L}$         & $N_\text{CP}^\text{W}$         & Number of samples of CP (144 and 16)\\ \bottomrule
\end{tabular}
\end{minipage}
}
\end{table}
To begin, important OFDM parameters are denoted in Table~\ref{Table:OFDM_Paramters} \cite{Guard_subcarrier, 80211}.

%
We define the $m$-th transmit OFDM symbol of an eNB and an AP including CP as
\begin{align}
\label{Eq:TX_OFDM_Symbol_LAA}
\!\!\!\!\! x_{m}^\text{L}(t)\! &=\!\! \sum_{k=0}^{N_{\text{FFT}}^\text{L}-1} \!\! a_{m,k}^\text{L} b(t; f_{k}^\text{L}, T_\text{CP}^\text{L}, T_\text{data}^\text{L}), 0\leq t < T_\text{total}^\text{L}, \\
\label{Eq:TX_OFDM_Symbol_WiFi}
x_{m}^\text{W}(t)\! &=\!\! \sum_{k=0}^{N_{\text{FFT}}^\text{W}-1} \!\! a_{m,k}^\text{W} b(t; f_{k}^\text{W}, T_\text{CP}^\text{W}, T_\text{data}^\text{W}), 0\leq t < T_\text{total}^\text{W},
\end{align}
where $a_{m,k}^\text{L}$ and $a_{m,k}^\text{W}$ are the message symbols of the $m$-th subcarrier of LAA-LTE and Wi-Fi, respectively, and $b(\cdot)$ is defined as
\begin{equation}
\!\!\!\!\!\!\!\!\!\!\!\!\!\!\!\!\!\!\!\!\!\!\!\!
b(t; f,T_1, T_2) \! = \! \left\{\begin{matrix}
\text{exp}\left( j 2\pi f (t+T_2 - T_1) \right ), & \!\!\!\! 0 \leq t < T_1,\\ 
\!\!\!\!\!\!\!\!\!\!\!\!\! \text{exp}\left( j 2\pi f (t - T_1) \right ), &\!\!\!\!\!\!\!\!\!\!\!\!\!\!\!\! T_1 \leq t < T_1 + T_2.
\end{matrix}\right. \!\!\!\!\!\!\!\!\!\!\!\!\!\!\!\!
\end{equation}
The definition of the function $b(\cdot)$ is illustrated in Fig.~\ref{Fig:b_function}.
The channel impulse response for the $k$-th path among $N_\text{tap}$ paths from the eNB to the $\alpha$-th AP and from the $\alpha$-th AP to the UE is denoted as $\nu_{k}^{\text{eNB},\text{AP}_\alpha}$ and $\nu_{k}^{\text{AP}_\alpha,\text{UE}}$, respectively.

Based on the notations, we firstly derive the received signals at an AP and UE, and then derive the impact of the difference in the physical-layer parameters on the effective channel. 

\subsection{Motivation of a New Derivation on the Effective Channels}
\label{sec:why_cannot_use_the_existing_channel_model}

To show the necessity of a new derivation on the effective channels, we start with discussing a transmission between homogeneous Wi-Fi nodes, i.e., the transmission from an AP to a STA.
The vector of the message signals in the $m$-th OFDM symbol are denoted by $\mathbf{a}_m^\text{W}\in \mathbb{C}^{N_\text{FFT}^\text{W} \times 1}$, and the IFFT and FFT matrices are defined by $\mathbf{Q}_\text{W} \in \mathbb{C}^{N_\text{FFT}^\text{W} \times N_\text{FFT}^\text{W}}$ and $\mathbf{F}_\text{W}\in \mathbb{C}^{N_\text{FFT}^\text{W} \times N_\text{FFT}^\text{W}}$ , respectively.
In addition, $\tilde{\mathbf{H}}_\text{AP,STA} \in \mathbb{C}^{N_\text{FFT}^\text{W} \times N_\text{FFT}^\text{W}}$ is the time-domain circulant channel matrix from the AP to the STA after the CP removal.
Hence, the frequency-domain response is as
\begin{equation}
\mathbf{F}_\text{W} \tilde{\mathbf{H}}_\text{AP,STA} \mathbf{Q}_\text{W} \times \mathbf{a}_m^\text{W} = \mathbf{H}_\text{AP,STA} \times \mathbf{a}_m^\text{W}
\end{equation}
where $\mathbf{H}_\text{AP,STA}=\mathbf{F}_\text{W} \tilde{\mathbf{H}}_\text{AP,STA} \mathbf{Q}_\text{W}$. 
In general, the effective channel matrix $\mathbf{H}_\text{AP,STA}$ is diagonal since $\tilde{\mathbf{H}}_\text{AP,STA}$ is circulant and $\mathbf{F}_\text{W}=\mathbf{Q}_\text{W}^\text{H}$.
Hence, each subcarrier has no effect on any other subcarriers due to the orthogonality between the subcarriers in case of the transmission between homogeneous nodes.

In the scenario of LAA-WLAN coexistence, however, the subcarrier spacing of LAA-LTE and Wi-Fi is different as in Table~\ref{Table:OFDM_Paramters}.
Since $\vartriangle\!\!f_\text{W}$ is larger than $\vartriangle\!\!f_\text{L}$, a Wi-Fi's subcarrier can overlap with around $\lceil 312.5/15 \rceil=21$ LAA-LTE's subcarriers.
According to the overlap, we can intuitively recognize that a Wi-Fi's subcarrier interferes multiple LAA-LTE's subcarriers and vice versa.
Therefore, the existing effective channel model between homogeneous nodes cannot be used in the LAA-WLAN coexistence scenario.

\subsection{Effective Channel Model from an AP to a UE}
\label{Sec:Channel Model from an AP to an UE}
\begin{figure}[!t]
\centering
\subfigure[$M'=2$ ($\tau_1>3.3 \mu s$)]{
\includegraphics[width=1\columnwidth]{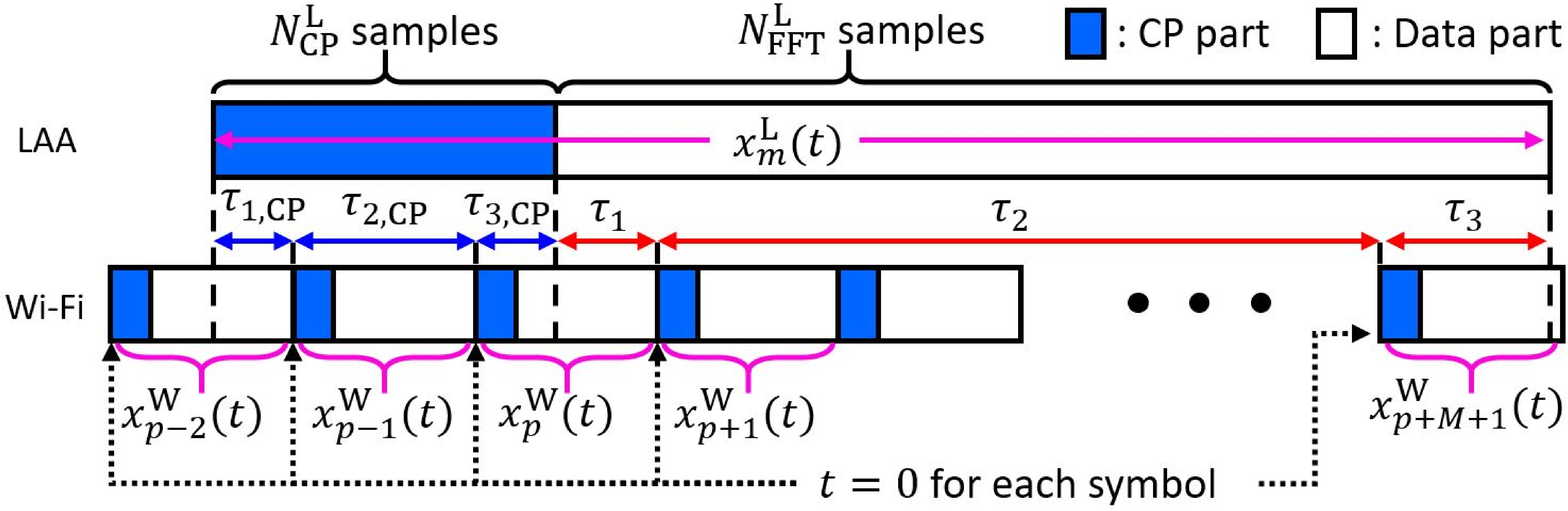}
\label{Fig:Received_Signal_at_UE_1}
}
\subfigure[$M'=1$ ($0 \leq \tau_1 \leq 3.3 \mu s$)]{
\includegraphics[width=1\columnwidth]{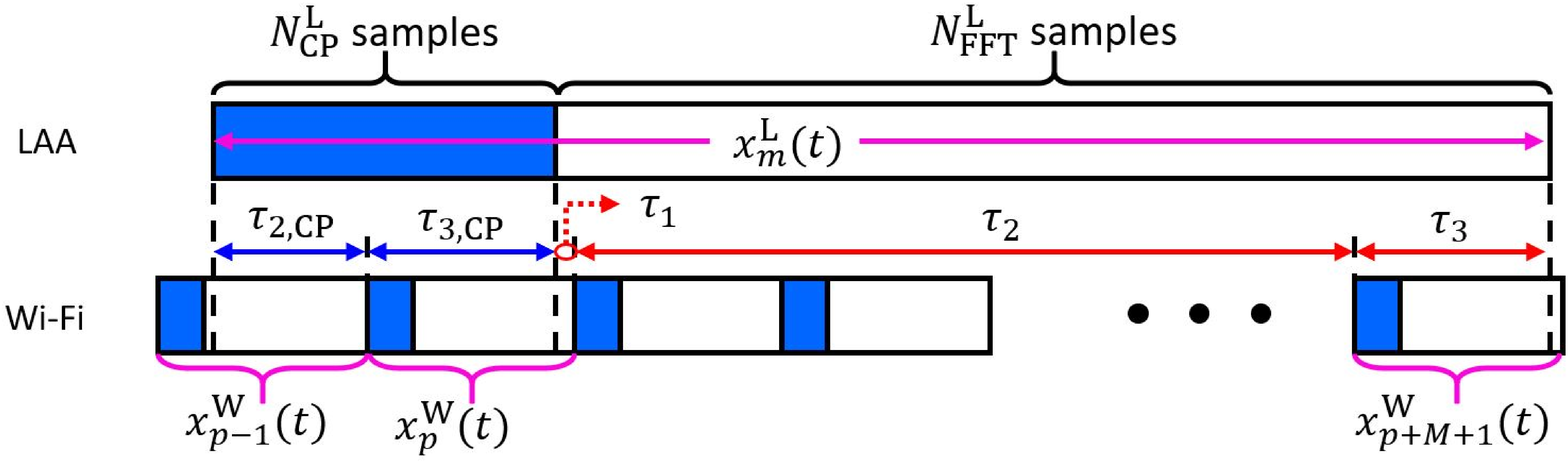}
\label{Fig:Received_Signal_at_UE_2}
}
\caption{Received signal at an UE for each $\tau_1$ case. The blue box means the cyclic prefix and the white box is the data duration.}
\label{Fig:Received_Signal_at_UE}
\vspace{-0.35in}
\end{figure}

To derive the effective channel model from an AP to a UE, we derive a Wi-Fi's signal received at the UE for the duration of one LAA-LTE's OFDM symbol. 
We start with the following notations: 
In Fig.~\ref{Fig:Received_Signal_at_UE},  
$\tau_1$ is time difference from the starting point of the data part of $x_m^\text{L}(t)$ to that of the CP of $x_{p+1}^\text{W}(t)$, where $0\leq \tau_1 < T_\text{total}^\text{W}$; 
$\tau_2$ is the duration from the starting point of the CP of $x_{p+1}^\text{W}(t)$ to that of the CP of $x_{p+M+1}^\text{W}(t)$, where $M$ is the number of Wi-Fi symbols completely overlapped with the data part of $x_m^\text{L}(t)$; 
$\tau_3$ is the duration of the part of $x_{p+M+1}^\text{W}$ overlapped with the data part of $x_m^\text{L}(t)$; $\tau_{1,\text{CP}}$, $\tau_{2,\text{CP}}$, and $\tau_{3,\text{CP}}$ denote the duration of the part of $x_{p-2}^\text{W}(t)$, $x_{p-1}^\text{W}(t)$, and $x_{p}^\text{W}(t)$ overlapped with the CP of $x_m^\text{L}(t)$, respectively; $\tau_{1,\text{CP}} = T_\text{CP}^\text{L}-(T_\text{total}^\text{W}-\tau_1)-T_\text{total}^\text{W}$ and $\tau_{3,\textrm{CP}} = T_\text{total}^\text{W}-\tau_1$; 
$\tau_\text{2,CP}$ is $T_\text{total}^\text{W}$ when $\tau_1>3.3 \mu s$ or $T_\text{CP}^\text{L}-(T_\text{total}^\text{W}-\tau_1)$ when $0 \leq \tau_1 \leq 3.3 \mu s$.

We first perform a case study for the variable $\tau_1$, where $\tau_2$, $\tau_3$, $\tau_{1,\text{CP}}$, $\tau_{2,\text{CP}}$, and $\tau_{3,\text{CP}}$ are functions of $\tau_1$, and then derive the throughput treating $\tau_1$ as a random variable. In addition, for notational simplicity, the AP index is not considered if not essential based on the analogy of the system model for every AP. 

%

For clarifying possible cases, we denote by $M'$ the number of Wi-Fi's symbols within the CP of $x_m^\text{L}(t)$ except for $x_p^\text{W}$ as in Fig.~\ref{Fig:Received_Signal_at_UE}. Then, we have 
$M' = \left \lceil {(T_\text{CP}^\text{L}-(T_\text{total}^\text{W}-\tau_1))/T_\text{total}^\text{W}} \right \rceil$.
Based on the definition of $M'$ and the range of $\tau_1$, and  from Table~\ref{Table:OFDM_Paramters}, we can obtain two possible cases, $M'=2$ ($\tau_1>3.3 \mu s$) and $M'=1$ ($0 \leq \tau_1 \leq 3.3 \mu s$).
We shall derive the effective channel only for the case of $M'=2$ which is a more general case. The derivation for the case of $M'=1$ can be readily obtained following the analogous derivation for the case of $M'=2$. 


\subsubsection{Samples for the CP Duration}
For $M'=2$, two symbols $x_{p-1}^\text{W}(t)$ and $x_{p-2}^\text{W}(t)$ are overlapped within the CP duration of the LAA-LTE's symbol $x_m^\text{L}(t)$ as in Fig.~\ref{Fig:Received_Signal_at_UE_1}.
For the time of $[0, \tau_{1,\text{CP}})$, the sample index is obtained by
\begin{equation}
\label{Eq:range_n_at_UE_case_1}
\hspace{.01in}
0 \leq \! \frac{n T_\text{CP}^\text{L}}{N_\text{CP}^\text{L}} \! < \tau_{1,\text{CP}}
\Leftrightarrow 0 \leq n <\!\! \left \lceil \frac{N_\text{CP}^\text{L}\tau_{1,\text{CP}}}{T_\text{CP}^\text{L}} \right \rceil \!\! -1 \triangleq \underline{N}_{\tau_1,\text{CP}}.
\end{equation}
%
%
Based on (\ref{Eq:range_n_at_UE_case_1}), the sampled points of $x_{p-2}^\text{W}(t)$ are 
\begin{align}
\!\!\!s_m^\text{W}& [n] \!= x_{p-2}^\text{W}(t)\big|_{t={n\cdot T_\text{CP}^\text{L}/N_\text{CP}^\text{L}}+T_\text{total}^\text{W}-\tau_{1,\text{CP}}}\\
\label{Eq:Received_Signal_at_UE_M_2_CP}&=\!\!\!\!\sum_{k=0}^{N_\text{FFT}^\text{W}-1} \!\!\!\! a_{p-2,k}^\text{W} 
b\!\left(\!\frac{n T_\text{CP}^\text{L}}{N_\text{CP}^\text{L}} + T_\text{total}^\text{W}- \tau_{1,\text{CP}}; f_k^\text{W}, T_\text{CP}^\text{W}, T_\text{data}^\text{W}\! \right) \!.\!\!
\end{align}
For (\ref{Eq:Received_Signal_at_UE_M_2_CP}), the vector form is written as
\begin{equation}
\mathbf{s}_{\tau_{1,\text{CP}}}^\text{W} \triangleq \mathbf{G}_{1,\text{CP}} \times \mathbf{a}_{p-2}^\text{W},
\end{equation}
where $\mathbf{s}_{\tau_{1,\text{CP}}}^\text{W} = [s_m^\text{W}[0],\ldots, s_m^\text{W}[\underline{N}_{\tau_1,\text{CP}}]]^\text{T}$ and $\mathbf{a}_{p-2}^\text{W} = [a_{p-2,0}^\text{W}, \ldots, a_{p-2,N_\text{FFT}^\text{W}-1}^\text{W}]^\text{T}$.
Here, the matrix $\mathbf{G}_{1,\text{CP}}$ is defined by
\begin{equation}
\![\mathbf{G}_\text{1,CP}]_{(n+1,k+1)} \! = \! b \! \left(\!\frac{n T_\text{CP}^\text{L}}{N_\text{CP}^\text{L}} \!+ T_\text{total}^\text{W}- \tau_{1,\text{CP}}; f_k^\text{W}\!, T_\text{CP}^\text{W}, T_\text{data}^\text{W} \!\right)\!, \!\!\!\!\!\!
\end{equation}
for $n\in[0,\underline{N}_{\tau_1,\text{CP}}]$ and $k\in[0,N_\text{FFT}^\text{W}-1]$, where $[\mathbf{X}]_{(i,j)}$ denotes the $(i,j)$-th element of the matrix $\mathbf{X}$.
%
%

For the period $\tau_\text{2,CP}$, the sample index can be obtained as
\begin{equation}
\left \lceil \frac{N_\text{CP}^\text{L}\tau_\text{1,CP}}{T_\text{CP}^\text{L}} \right \rceil \!
 \leq n < \!\!
\left \lceil \frac{N_\text{CP}^\text{L}(\tau_\text{1,CP}+\tau_\text{2,CP})}{T_\text{CP}^\text{L}} \right \rceil \! -1
\! \triangleq \! \underline{N}_{\tau_2, \text{CP}}, \!\!\!\!
\end{equation}
and the sampled points of $x_{p-1}^\text{W}(t)$ are 
\begin{align}
\!\!\!s_m^\text{W} [n] \!&= x_{p-1}^\text{W}(t)\big|_{t={nT_\text{CP}^\text{L}/N_\text{CP}^\text{L} - \tau_{1,\text{CP}}}}\\
&=\!\!\!\!\sum_{k=0}^{N_\text{FFT}^\text{W}-1} \!\!\!\! a_{p-1,k}^\text{W} 
b\!\left(\!\frac{nT_\text{CP}^\text{L}}{N_\text{CP}^\text{L}} - \tau_{1,\text{CP}}; f_k^\text{W}, T_\text{CP}^\text{W}, T_\text{data}^\text{W}\! \right) \!,\!\!
\end{align}
where the time variable $t$ for each Wi-Fi symbol is reset to 0 at the beginning of each Wi-Fi symbol for notational simplicity, as defined in \eqref{Eq:TX_OFDM_Symbol_WiFi}  and shown in Fig. \ref{Fig:Received_Signal_at_UE_1}. 
Thus, the vector of the sampled points of $x_{p-1}^\text{W}(t)$ is derived as 
\begin{equation}
\mathbf{s}_{\tau_{2,\text{CP}}}^\text{W} \triangleq \mathbf{G}_{2,\text{CP}} \times \mathbf{a}_{p-1}^\text{W},
\end{equation}
where $\mathbf{s}_{\tau_{2,\text{CP}}}^\text{W} = [s_m^\text{W}[\underline{N}_{\tau_1,\text{CP}}+1],\ldots, s_m^\text{W}[\underline{N}_{\tau_2,\text{CP}}]]^\text{T}$ and
 $\mathbf{a}_{p-1}^\text{W} = [a_{p-1,0}^\text{W} ,\ldots, a_{p-1,N_\text{FFT}^\text{W}-1}^\text{W}]^\text{T}$.
Here, the matrix $\mathbf{G}_\text{2,CP}$ is defined by
\begin{equation}
[\mathbf{G}_\text{2,CP}]_{(n+1,k+1)} \! = \! b \! \left(\frac{n T_\text{CP}^\text{L}}{N_\text{CP}^\text{L}} - \tau_{1,\text{CP}}; f_k^\text{W}, T_\text{CP}^\text{W}, T_\text{data}^\text{W} \right)\!, 
\end{equation}
where $n\in[\underline{N}_{\tau_1,\text{CP}}+1,\underline{N}_{\tau_2,\text{CP}}]$ and $k\in[0,N_\text{FFT}^\text{W}-1]$.
Finally, for the period $\tau_\text{3,CP}$, the sample index is calculated by
\begin{equation}
\!
\underline{N}_{\tau_2, \text{CP}} + 1
 \leq n < \!\!
\left \lceil \frac{N_\text{CP}^\text{L}(\tau_\text{1,CP}+\tau_\text{2,CP}+\tau_\text{3,CP})}{T_\text{CP}^\text{L}} \right \rceil \! -1
\! \triangleq \! \underline{N}_{\tau_3, \text{CP}}, \!\!\!\!
\end{equation}
and the vector of the sampled points from $x_{p}^\text{W}(t)$ is derived as 
\begin{equation}
\mathbf{s}_{\tau_{3,\text{CP}}}^\text{W} \triangleq \mathbf{G}_{3,\text{CP}} \times \mathbf{a}_{p}^\text{W},
\end{equation}
where $\mathbf{s}_{\tau_{3,\text{CP}}}^\text{W} = [s_m^\text{W}[\underline{N}_{\tau_2,\text{CP}}+1],\ldots, s_m^\text{W}[\underline{N}_{\tau_3,\text{CP}}]]^\text{T}$ and
 $\mathbf{a}_{p}^\text{W} = [a_{p,0}^\text{W} ,\ldots, a_{p,N_\text{FFT}^\text{W}-1}^\text{W}]^\text{T}$.
The matrix $\mathbf{G}_\text{3,CP}$ is defined by
\begin{equation}
[\mathbf{G}_\text{3,CP}]_{(n+1,k+1)} \! = 
\! b \! \left(\frac{n T_\text{CP}^\text{L}}{N_\text{CP}^\text{L}} - \tau_{1,\text{CP}} - \tau_{2,\text{CP}}; f_k^\text{W}, T_\text{CP}^\text{W}, T_\text{data}^\text{W} \right)\!, 
\end{equation}
for $n\in[\underline{N}_{\tau_2,\text{CP}}+1,N_\text{CP}^\text{L}]$ and $k\in[0,N_\text{FFT}^\text{W}-1]$.

\subsubsection{Samples for the Data Duration}
To ease notation, we reset the sample index for the data part due to the difference in the sampling frequencies for the CP and data in LAA-LTE.
Specifically, in Fig.~\ref{Fig:Received_Signal_at_UE_1}, the $p$-th Wi-Fi's symbol $x_p^\text{W}$ is overlapped with the data part of the LAA-LTE's symbol, and the sample index is obtained as
\begin{equation}
0 \leq \frac{n T_\text{data}^\text{L}}{N_\text{FFT}^\text{L}} < \tau_1
\Longleftrightarrow
0 \leq n < \left \lceil \frac{\tau_1 N_\text{FFT}^\text{L}}{T_\text{data}^\text{L}} \right \rceil - 1
\triangleq \underline{N}_1.
\end{equation}
For the duration $[0, \tau_1)$, the sampled points of $x_p^\text{W}(t)$ are written by
\begin{align}
\label{Eq:Received_Signal_at_UE_M_2_data}
\!\!\!\!\!s_m^\text{W}& [n] \!= x_{p}^\text{W}(t)\big|_{t={n T_\text{data}^\text{L}/N_\text{FFT}^\text{L}}+T_\text{total}^\text{W}-\tau_1}\\
&\!=\!\sum_{k=0}^{N_\text{FFT}^\text{W}-1} \!\! a_{p,k}^\text{W} 
b\left(\frac{n T_\text{data}^\text{L}}{N_\text{FFT}^\text{L}}+T_\text{total}^\text{W}-\tau_1; f_k^\text{W}, T_\text{CP}^\text{W}, T_\text{data}^\text{W} \right) \!,\!\!
\end{align}
and the vector form of the sampled points is derived as
\begin{equation}
\mathbf{s}_{\tau_{1}}^\text{W} \triangleq \mathbf{G}_{\tau_1} \times \mathbf{a}_{p}^\text{W},
\end{equation}
where $\mathbf{a}_{p}^\text{W} = [a_{p,0}^\text{W} ,\ldots, a_{p,N_\text{FFT}^\text{W}-1}^\text{W}]^\text{T}$
and $\mathbf{s}_{\tau_{1}}^\text{W} = [s_m^\text{W}[0] ~ ,\ldots, \\ s_m^\text{W}[\underline{N}_{1}]]^\text{T}$.
The matrix $\mathbf{G}_{\tau_1}$ is defined by
\begin{equation}
\![\mathbf{G}_{\tau_1}]_{(n+1,k+1)} \! = 
\! b\left(\frac{n T_\text{data}^\text{L}}{N_\text{FFT}^\text{L}}+T_\text{total}^\text{W}-\tau_1; f_k^\text{W}, T_\text{CP}^\text{W}, T_\text{data}^\text{W} \right)\!, 
\end{equation}
where $n\in[0,\underline{N}_1]$ and $k\in[0,N_\text{FFT}^\text{W}-1]$.

In the data part of the LAA-LTE's symbol, there are $M$ whole Wi-Fi's symbols, where $M=\lfloor(T_\text{data}^\text{L}-\tau_1)/T_\text{total}^\text{W} \rfloor$.
Hence, for the duration $[\tau_1 + T_\text{total}^\text{W}(q-1), \tau_1+T_\text{total}^\text{W}q)$, $q=1,\ldots,M$,\footnote{Since $\tau_2 = M \times T_\text{total}^\text{W}$, the time duration is $[\tau_1, \tau_2)$ for all $q$.}
the sample index $n$ should satisfy 
\begin{equation}
\label{Eq:Received_Signal_at_UE_M_2_data_tau_2}
\overline{N}_{\tau_2,q} \!\!
\leq  n < \underline{N}_{\tau_2,q},
\end{equation}
where 
\begin{align}
    \overline{N}_{\tau_2,q} =  \left \lceil  \frac{N_\text{FFT}^\text{L}(\tau_1+T_\text{total}^\text{W}(q-1))}{T_\text{data}^\text{L}}  \right \rceil,
\end{align}
\begin{align}
    \underline{N}_{\tau_2,q}=\left \lceil  \frac{N_\text{FFT}^\text{L}(\tau_1+T_\text{total}^\text{W}q)}{T_\text{data}^\text{L}}  \right \rceil - 1.
\end{align}
Hence, for the duration, the sampled points from $x_{p+q}^\text{W}(t)$ are 
\begin{align}
\!\!\!\!\!\!\!\!\!&s_m^\text{W}[n] \!= x_{p+q}^\text{W}\left (t \right)\big|_{t={n T_\text{data}^\text{L}/N_\text{FFT}^\text{L}}-(\tau_1+T_\text{total}^\text{W}(q-1))}\\
&=\!\!\!\!\!\sum_{k=0}^{N_\text{FFT}^\text{W}-1}  \!\!\!\!a_{p+q,k}^\text{W} 
b\!\left(\!\frac{n T_\text{data}^\text{L}}{N_\text{FFT}^\text{L}}-(\tau_1 \!+T_\text{total}^\text{W}(q-1)); f_k^\text{W}\!, T_\text{CP}^\text{W}, T_\text{data}^\text{W}\!\! \right) \!,\!\!\!\!\!
\end{align}
and the vector form of the sampled points is obtained as
\begin{equation}
\mathbf{s}_{\tau_{2},q}^\text{W} \triangleq \mathbf{G}_{\tau_2,q} \times \mathbf{a}_{p+q}^\text{W},
\end{equation}
where $\mathbf{s}_{\tau_{2},q}^\text{W} = [s_m^\text{W}[\overline{N}_{\tau_2,q}] ,\ldots,
 s_m^\text{W}[\underline{N}_{\tau_2,q}]]^\text{T}$ and $\mathbf{a}_{p+q}^\text{W} = [a_{p+q,0}^\text{W} ,\ldots, a_{p+q,N_\text{FFT}^\text{W}-1}^\text{W}]^\text{T}$.
The matrix $\mathbf{G}_{\tau_2,q}$ is defined by
\begin{align}
[\mathbf{G}_{\tau_2,q}&]_{(n-\overline{N}_{\tau_2,q}+1,k+1)} \! = \nonumber\\
&b\left(\frac{n T_\text{data}^\text{L}}{N_\text{FFT}^\text{L}}-(\tau_1+T_\text{total}^\text{W}(q-1)); f_k^\text{W}, T_\text{CP}^\text{W}, T_\text{data}^\text{W} \right)\!, 
\end{align}
where $n\in[\overline{N}_{\tau_2,q},\underline{N}_{\tau_2,q}]$ and $k\in[0,N_\text{FFT}^\text{W}-1]$.
Finally, for the duration $[\tau_1+\tau_2, T_\text{data}^\text{L})$, the sample index $n$ should satisfy 
\begin{equation}
    \left \lceil \frac{N_\text{FFT}^\text{W}(\tau_1+\tau_2)}{T_\text{data}^\text{L}} \right \rceil \triangleq \overline{N}_\text{last}
    \leq n < N_\text{FFT}^\text{L}-1,
\end{equation}
and the sampled points of $x_{p+M+1}^\text{W}$ are
\begin{align}
\!\!\!\!\!&s_m^\text{W}[n] \!= x_{p+M+1}^\text{W}\left (t \right)\big|_{t={n T_\text{data}^\text{L}/N_\text{FFT}^\text{L}}-(\tau_1+\tau_2)}\\
&\!=\!\sum_{k=0}^{N_\text{FFT}^\text{W}-1}  a_{p+q,k}^\text{W} 
b\left(\frac{n T_\text{data}^\text{L}}{N_\text{FFT}^\text{L}}-(\tau_1+\tau_2); f_k^\text{W}, T_\text{CP}^\text{W}, T_\text{data}^\text{W} \right) \!.\!\!
\end{align}
The vector form is derived as
\begin{equation}
    \mathbf{s}_{\tau_3}^\text{W} \triangleq \mathbf{G}_{\tau_3}\times \mathbf{a}_{p+M+1}^\text{W},
\end{equation}
where $\mathbf{s}_{\tau_3}^\text{W}=[s_m^\text{W}[\overline{N}_\text{last}] ,\ldots, s_m^\text{W}[N_\text{FFT}^\text{L}-1]]^\text{T}$ and $\mathbf{a}_{p+M+1}^\text{W}=[a_{p+M+1,0}^\text{W} ,\ldots, a_{p+M+1,N_\text{FFT}^\text{W}-1}^\text{W}]^\text{T}$.
The matrix $\mathbf{G}_{\tau_3}$ is defined by
\begin{equation}
    [\mathbf{G}_{\tau_3}]_{(n-\overline{N}_\text{last}+1,k+1)} \! = \! b \! \left( \! \frac{n T_\text{data}^\text{L}}{N_\text{FFT}^\text{L}}-(\tau_1+\tau_2); f_k^\text{W}, T_\text{CP}^\text{W}, T_\text{data}^\text{W} \! \right)\!,
\end{equation}
where $n\in[\overline{N}_\text{last},N_\text{FFT}^\text{L}-1]$ and $k\in[0,N_\text{FFT}^\text{W}-1]$.

\subsubsection{Frequency-Domain Effective Channel Matrix}
\begin{figure*}
\centering
\resizebox{1.0\linewidth}{!}{
\begin{minipage}{\linewidth}
\begin{align} 
\label{Eq:Received_Signal_at_UE_M_2_overall}
\mathbf{s}_{\textrm{W}}&=\begin{bmatrix}
\mathbf{s}_{\tau_{1,\textrm{CP}}}^\textrm{W}\\ 
\mathbf{s}_{\tau_{2,\textrm{CP}}}^\textrm{W}\\ 
\mathbf{s}_{\tau_{3,\textrm{CP}}}^\textrm{W}\\ 
\mathbf{s}_{\tau_{1}}^\textrm{W}\\ 
\mathbf{s}_{\tau_{2,1}}^\textrm{W}\\ 
\vdots\\ 
\mathbf{s}_{\tau_{2,M}}^\textrm{W}\\
\mathbf{s}_{\tau_{3}}^\textrm{W}
\end{bmatrix}
=\underbrace{\begin{bmatrix}
\mathbf{G}_{1,\textrm{CP}} &  &  &  &  &  &  & \\ 
 & \mathbf{G}_{2,\textrm{CP}}  &  &  &  &  &  & \\ 
 &  & \mathbf{G}_{3,\textrm{CP}} &  &  &  &  & \\ 
 &  &  & \mathbf{G}_{\tau_{1}} &  &  &  & \\ 
 &  &  &  & \mathbf{G}_{\tau_{\tau_{2,1}}} &  &  & \\ 
 &  &  &  &  & \ddots &  & \\ 
 &  &  &  &  &  & \mathbf{G}_{\tau_{\tau_{2,M}}}  & \\ 
 &  &  &  &  &  &  & \mathbf{G}_{\tau_{\tau_{3}}} 
\end{bmatrix}}_{\triangleq \mathbf{K}_{\textrm{AP},\textrm{UE}}}
\underbrace{\begin{bmatrix}
\mathbf{a}_{p-2}^\textrm{W}\\ 
\mathbf{a}_{p-1}^\textrm{W}\\ 
\mathbf{a}_{p}^\textrm{W}\\ 
\mathbf{a}_{p}^\textrm{W}\\ 
\mathbf{a}_{p+1}^\textrm{W}\\ 
\vdots\\ 
\mathbf{a}_{p+M}^\textrm{W}\\ 
\mathbf{a}_{p+M+1}^\textrm{W}
\end{bmatrix}}_{\triangleq \tilde{\mathbf{a}}_{\textrm{W}}}
\end{align}
\noindent\makebox[\linewidth]{\rule{16.5cm}{0.4pt}}
\end{minipage}
}
\end{figure*}
By combining the derived sampled points for durations, the discrete-time Wi-Fi transmit vector for the overall duration of $T_\text{CP}^\text{L}+T_\text{data}^\text{L}$ is derived as 
\begin{equation}
\mathbf{s}_{\textrm{W}}=\mathbf{K}_{\textrm{AP},\textrm{UE}} \times \tilde{\mathbf{a}}_{\textrm{W}},
\end{equation}
each element of which is constructed as~(\ref{Eq:Received_Signal_at_UE_M_2_overall}).

At this point, to consider the effect of the channel impulses on the received signal, we denote the discrete-time-domain channel matrix from the $\alpha$-th AP to the UE as $\tilde{\mathbf{H}}_{\textrm{AP}_\alpha,\textrm{UE}}\in \mathbb{C}^{N_\text{FFT}^\text{L} \times N_\text{total}^\text{L}}$, which is defined by 
\begin{align}
\label{Eq:Channel_Impulse_Matrix_for_AP_to_UE}
&\tilde{\mathbf{H}}_{\textrm{AP}_\alpha,\textrm{UE}}\nonumber\\
&=
\begin{bmatrix}
\overbrace{0 \cdots 0}^{N_\text{CP}^\text{L}-N_\text{tap}+1} \!\!\!\!\!\! &\nu_{N_\text{tap}}^{\text{AP}_\alpha,\text{UE}} & \cdots & \nu_1^{\text{AP}_\alpha,\text{UE}} &  &  \\ 
&  & \ddots & \ddots & \ddots & \\ 
&  &  & \nu_{N_\text{tap}}^{\text{AP}_\alpha,\text{UE}} & \cdots & \nu_1^{\text{AP}_\alpha,\text{UE}}
\end{bmatrix}\!.
\end{align}

Then, the received interference signal at the UE due to the AP's signal is derived by
\begin{align}
\mathbf{r}_\text{W} &= \tilde{\mathbf{H}}_{\text{AP}_{\alpha},\text{UE}}  \times \mathbf{s}_\text{W}
\label{Eq:Received_Signal_at_UE_with_Channel}
=\tilde{\mathbf{H}}_{\text{AP}_{\alpha},\text{UE}}  \mathbf{K}_\text{AP,UE} \times \tilde{\mathbf{a}}_\text{W}.
\end{align}
By multiplying both sides of (\ref{Eq:Received_Signal_at_UE_with_Channel}) with the FFT matrix of LAA-LTE $\mathbf{F}_\text{L} \in \mathbb{C}^{N_\text{FFT}^\text{L}\times N_\text{FFT}^\text{L}}$, we can obtain the frequency-domain interference at the UE as
\begin{equation}
\mathbf{F}_\text{L} \times \mathbf{r}_\text{L} =\mathbf{F}_\text{L} \tilde{\mathbf{H}}_{\text{AP}_{\alpha},\text{UE}} \mathbf{K}_\text{AP,UE} \times \tilde{\mathbf{a}}_\text{W}.
\end{equation}
As a result, the frequency-domain effective channel matrix from the $\alpha$-th AP to the UE can be defined as 
\begin{equation}
\mathbf{H}_{\text{AP}_{\alpha},\text{UE}} =\mathbf{F}_\text{L} \tilde{\mathbf{H}}_{\text{AP}_{\alpha},\text{UE}}  \mathbf{K}_\text{AP,UE}.
\label{eq:channel_AP_UE_final}
\end{equation}

\subsection{Effective Channel Model from an eNB to an AP}
\label{Sec:Channel Model from an eNB to a STA}
\begin{figure}[!t]
\centering
\subfigure[$0\leq \tau < T_\text{total}^\text{L} - T_\text{total}^\text{W}$]{
\includegraphics[width=1\columnwidth]{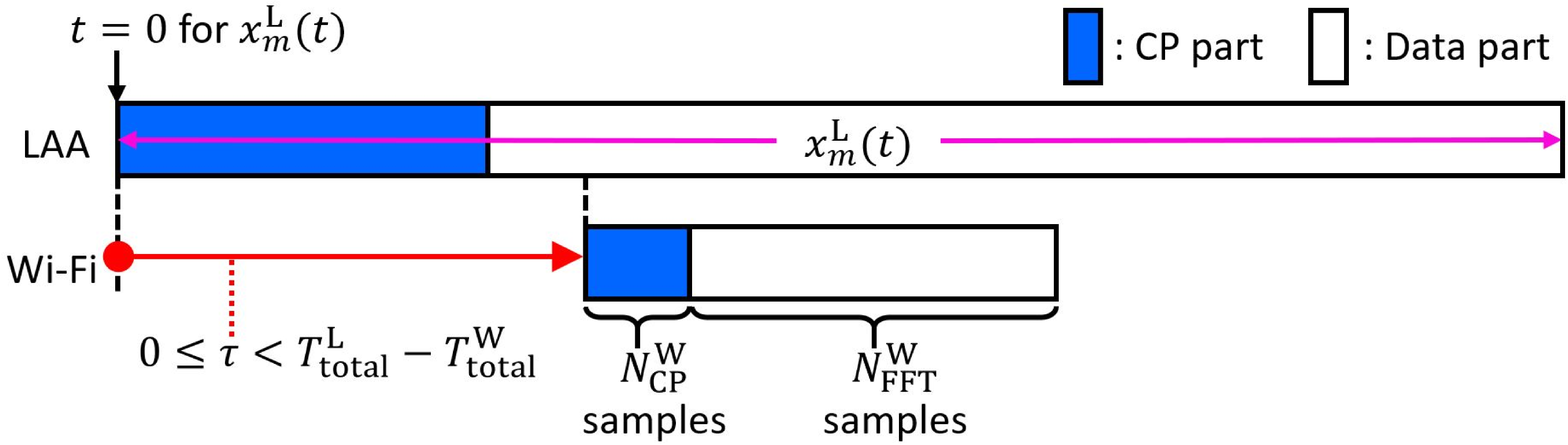}
\label{Fig:Received_Signal_at_STA_1}
}
\subfigure[$-T_\text{total}^\text{W} \leq \tau < 0$]{
\includegraphics[width=1\columnwidth]{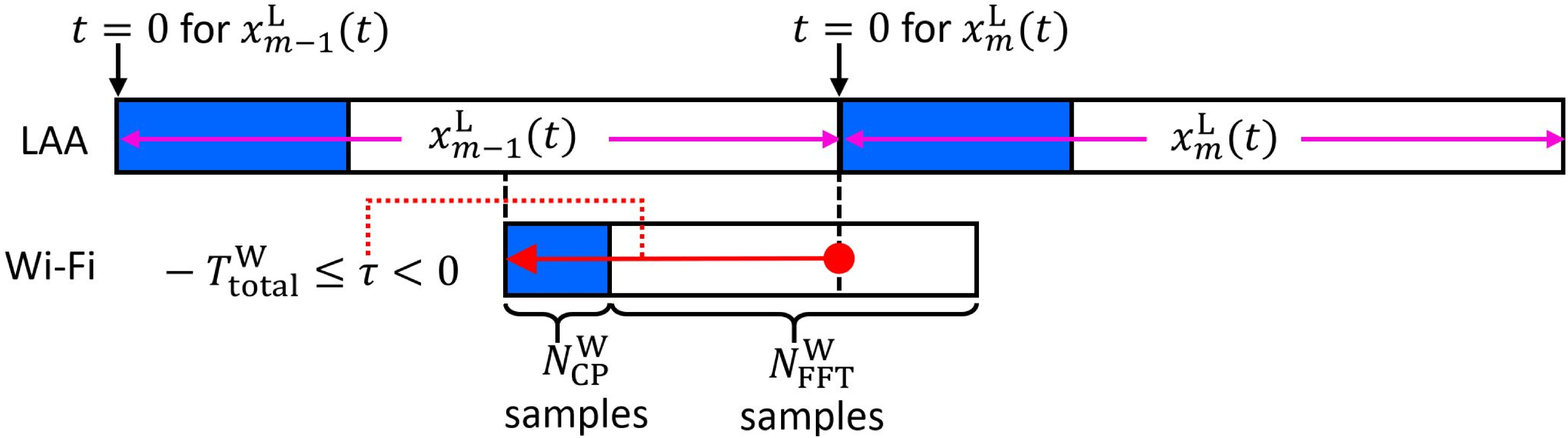}
\label{Fig:Received_Signal_at_STA_2}
}
\caption{Received signal at an AP for each $\tau$ case. The blue box means the cyclic prefix, and the white box is the data.}
\label{Fig:Received_Signal_at_STA}
\end{figure}

To derive the eNB's signal received at an AP, $\tau$ is defined as the time difference between the starting point of an LAA-LTE's symbol and that of the considered Wi-Fi's symbol, as shown in Fig.~\ref{Fig:Received_Signal_at_STA}. We first perform a case study for the variable $\tau$, and then derive the throughput treating it as a random variable. 
For the frequency-domain CCA, an AP senses signals for the Wi-Fi's OFDM symbol duration. Thus, we derive the LAA-LTE's signal received at an AP for the duration of one Wi-Fi's OFDM symbol. 

There are two cases depending on how a Wi-Fi's OFDM symbol duration overlaps with multiple LAA-LTE's OFDM symbols as follows:
\begin{itemize}
\item {Case 1 ($0\!\!\leq\! \tau \!<\!\! T_\text{total}^\text{L} \!-\! T_\text{total}^\text{W}$)}: a Wi-Fi's symbol duration is overlapped with only one LAA-LTE's symbol as in Fig.~\ref{Fig:Received_Signal_at_STA_1}.
\item {Case 2 ($-T_\text{total}^\text{W} \leq \tau < 0$)}: two LAA-LTE's symbols are overlapped with a Wi-Fi's symbol duration as in Fig.~\ref{Fig:Received_Signal_at_STA_2}.
\end{itemize}

\subsubsection{Case 1}
In Fig.~\ref{Fig:Received_Signal_at_STA_1}, an AP  obtains $N_\text{CP}^\text{W}$ samples and $N_\text{FFT}^\text{W}$ samples from $x_m^\text{L}(t)$ for the CP duration and for the data duration, respectively. 
With the sample index $n$, the sampled points from $x_m^\text{L}(t)$ are written as
\begin{align}
s_\text{L}[n] &= x_{m}^\text{L}(t)\big|_{t={n\cdot T_\text{total}^\text{W}/N_\text{total}^\text{W}}+\tau}, ~~n\in[0,N_\text{total}^\text{W}-1]\\
\label{Eq:Received_Signal_at_STA_1}
&=\sum_{k=0}^{N_\text{FFT}^\text{L}-1} \!\! a_{m,k}^\text{L} b\left(\frac{n\cdot T_\text{total}^\text{W}}{N_\text{total}^\text{W}} +\tau; f_k^\text{L}, T_\text{CP}^\text{L}, T_\text{data}^\text{L} \right),
\end{align}
where $N_\text{total}^\text{W}=N_\text{CP}^\text{W}+N_\text{FFT}^\text{W}$.
The vector form of (\ref{Eq:Received_Signal_at_STA_1}) is defined as
\begin{equation}
\mathbf{s}_\text{L} \triangleq \mathbf{K}_\text{eNB,AP} \times \mathbf{a}_\text{L},
\end{equation} 
where $\mathbf{a}_\text{L}\!\! = \! [a_{m,0}^\text{L},\ldots, a_{m,N_\text{FFT}^\text{L}-1}^\text{L} ]^\text{T}$ and $\mathbf{s}_\text{L} \!\! = \! [ s_\text{L}[0],\ldots, \\ s_\text{L}[N_\text{total}^\text{L}-1] ]^\text{T}$, where $N_\text{total}^\text{L}=N_\text{CP}^\text{L}+N_\text{FFT}^\text{L}$.
The matrix $\mathbf{K}_\text{eNB,AP}$ is defined as
\begin{equation}
\!
\left [ \mathbf{K}_\text{eNB,AP} \right ]_{(n+1, k+1)} =  b\left(\frac{n\cdot T_\text{total}^\text{W}}{N_\text{total}^\text{W}} +\tau; f_k^\text{L}, T_\text{CP}^\text{L}, T_\text{data}^\text{L} \right),
\end{equation}
for $n\in[0,N_\text{total}^\text{W}-1]$ and $k \in [0, N_\text{FFT}^\text{L}-1]$.

\subsubsection{Case 2}
In Fig.~\ref{Fig:Received_Signal_at_STA_2}, the $(m-1)$-th LAA-LTE's symbol is overlapped for the index
\begin{equation}
0 \leq \frac{n T_\text{total}^\text{W}}{N_\text{total}^\text{W}} < \left | \tau \right | \Longleftrightarrow
n = 0, \ldots, \left \lceil \frac{\left | \tau \right | N_\text{total}^\text{W}}{T_\text{total}^\text{W}} \right \rceil \!-\!1
\triangleq \underline{N}_\tau,
\end{equation}
while the $m$-th LAA-LTE's symbol is overlapped for the sample index
\begin{equation}
\left | \tau \right | \leq \frac{n \cdot T_\text{total}^\text{W}}{N_\text{total}^\text{W}} < T_\text{total}^\text{W} \\
\Longleftrightarrow n = \underline{N}_\tau + 1, \ldots , N_\text{total}^\text{W}.
\end{equation}
Thus, the $N_\text{total}^\text{W}$ samples of the LAA-LTE's symbols are obtained as follows.
For $n\in[0,\underline{N}_\tau]$, $s_\text{L}[n]$ is derived as
\begin{align}
\!\!\!\!
s_\text{L}&[n] = x_{m-1}^\text{L}(t)\big|_{t={n \cdot T_\text{total}^\text{W}/N_\text{total}^\text{W}}+T_\text{total}^\text{L}-\left | \tau \right |}\\
\label{Eq:Received_Signal_at_STA_2_1}
&=\!\!\!\sum_{k=0}^{N_\text{FFT}^\text{L}-1} \!\!\!\!a_{m-1,k}^\text{L} b\!\left(\!\frac{n T_\text{total}^\text{W}}{N_\text{total}^\text{W}} + T_\text{total}^\text{L}\!-\!\left | \tau \right |; f_k^\text{L}, T_\text{CP}^\text{L}, T_\text{data}^\text{L}\! \right)\!.
\end{align}
For $n \in [\underline{N}_\tau+1,N_\text{total}^\text{W}-1]$, we have
\begin{align}
s_\text{L}[n] &= x_{m}^\text{L}(t)\big|_{t={n \cdot T_\text{total}^\text{W}/N_\text{total}^\text{W}}-\left | \tau \right |} \\
\label{Eq:Received_Signal_at_STA_2_2}
&=\sum_{k=0}^{N_\text{FFT}^\text{L}-1} a_{m,k}^\text{L} b\left(\frac{n\cdot T_\text{total}^\text{W}}{N_\text{total}^\text{W}} -\left | \tau \right |; f_k^\text{L}, T_\text{CP}^\text{L}, T_\text{data}^\text{L} \right).
\end{align}
In the vector form with (\ref{Eq:Received_Signal_at_STA_2_1}) and (\ref{Eq:Received_Signal_at_STA_2_2}), we have
\begin{equation}
\label{Eq:K_matrix_at_STA_2}
\mathbf{s}_\text{L} \triangleq \mathbf{K}_\text{eNB,AP} \times \mathbf{a}_\text{L}=
\begin{bmatrix}
\mathbf{K}_{\tau} & \\ 
 &\mathbf{K}_{1-\tau} 
\end{bmatrix}
\mathbf{a}_\text{L},
\end{equation} 
where $\mathbf{s}_\text{L}=[s_\text{L}[0],\ldots, s_\text{L}[\underline{N}_\tau]~ s_\text{L}[\underline{N}_\tau + 1] ,\ldots,s_\text{L}[N_\text{total}^\text{W}-1]]^\text{T}$ and 
$\mathbf{a}_\text{L} = [a_{m-1,0}^\text{L},\ldots, a_{m-1,N_\text{FFT}^\text{L}-1}^\text{L}, \\
a_{m,0}^\text{L},\ldots, a_{m,N_\text{FFT}^\text{L}-1}^\text{L}]^\text{T}$.
In (\ref{Eq:K_matrix_at_STA_2}), $\mathbf{K}_{\tau}$ and 
 $\mathbf{K}_{1-\tau}$ are defined as follows:
\begin{equation}
\!\left [ \mathbf{K}_\tau \right ]_{(n+1,k+1)} \!=\! b\!\left(\frac{n\!\cdot\! T_\text{total}^\text{W}}{N_\text{total}^\text{W}} + T_\text{total}^\text{L} \!-\!\left | \tau \right |; f_k^\text{L}, T_\text{CP}^\text{L}, T_\text{data}^\text{L} \! \right)\!\!
\end{equation}
for $n\in [0,\underline{N}_\tau]$ and $k\in[0,N_\text{FFT}^\text{L}]$, and 
\begin{equation}
\!\left [ \mathbf{K}_{1-\tau} \right ]_{({n-\underline{N}_\tau},k+1)} = b\left(\frac{n\!\cdot\! T_\text{total}^\text{W}}{N_\text{total}^\text{W}} -\left | \tau \right |; f_k^\text{L}, T_\text{CP}^\text{L}, T_\text{data}^\text{L} \right)\!\!
\end{equation}
for $n\in [\underline{N}_\tau + 1, N_\text{FFT}^\text{W}]$ and $k\in[0,N_\text{FFT}^\text{L}]$.

\subsubsection{Frequency-Domain Effective Channel Matrix}
To take into account the effect of the channel impulses on the received signal, we denote the discrete-time-domain channel matrix from the eNB to the $\alpha$-th AP as $\tilde{\mathbf{H}}_{\textrm{eNB},\textrm{AP}_\alpha}\in \mathbb{C}^{N_\text{FFT}^\text{W} \times N_\text{total}^\text{W}}$, where 
\begin{align}
\label{Eq:Channel_Impulse_Matrix_for_eNB_to_STA}
&\tilde{\mathbf{H}}_{\textrm{eNB},\textrm{AP}_\alpha}\nonumber\\
&=
\begin{bmatrix}
\overbrace{0 \cdots 0}^{N_\text{CP}^\text{W}-N_\text{tap}+1} \!\!\!\!\!\! &\nu_{N_\text{tap}}^{\text{eNB},\text{AP}_\alpha} & \cdots & \nu_1^{\text{eNB},\text{AP}_\alpha} &  &  \\ 
&  & \ddots & \ddots & \ddots & \\ 
&  &  & \nu_{N_\text{tap}}^{\text{eNB},\text{AP}_\alpha} & \cdots & \nu_1^{\text{eNB},\text{AP}_\alpha}
\end{bmatrix}\!.
\end{align}
Then, the eNB's signal received at the $\alpha$-th AP is expressed by
\begin{align}
\mathbf{r}_\text{L} &= \tilde{\mathbf{H}}_{\text{eNB},\text{AP}_\alpha} \times \mathbf{s}_\text{L} 
\label{Eq:Received_Signal_at_STA_with_Channel}
=\tilde{\mathbf{H}}_{\text{eNB},\text{AP}_\alpha}  \mathbf{K}_{\text{eNB},\text{AP}} \times \mathbf{a}_\text{L}.
\end{align}
By multiplying both sides of (\ref{Eq:Received_Signal_at_STA_with_Channel}) with $\mathbf{F}_\text{W} \in \mathbb{C}^{N_\text{FFT}^\text{W}\times N_\text{FFT}^\text{W}}$, the FFT matrix of Wi-Fi, we obtain the frequency-domain received signal at the AP as
\begin{equation}
\mathbf{F}_\text{W} \times \mathbf{r}_\text{L} =\mathbf{F}_\text{W} \tilde{\mathbf{H}}_{\text{eNB},\text{AP}_\alpha}  \mathbf{K}_\text{eNB,AP} \times \mathbf{a}_\text{L}.
\end{equation}
As a result, in the frequency domain, the effective channel matrix from the eNB to the $\alpha$-th AP can be defined as follows.
\begin{equation}
\mathbf{H}_{\text{eNB},\text{AP}_\alpha} =\mathbf{F}_\text{W} \tilde{\mathbf{H}}_{\text{eNB},\text{AP}_\alpha} \mathbf{K}_\text{eNB,AP}.
\label{eq:channel_eNB_AP_final}
\end{equation}

\subsubsection{Results of the Effective Channel Matrices}
%
%
\begin{figure}[!t]
\centering
\subfigure[Averaged squared absolute values of $\mathbf{H}_{\text{eNB},\text{AP}_\alpha}$]{
\includegraphics[width=0.97\columnwidth]{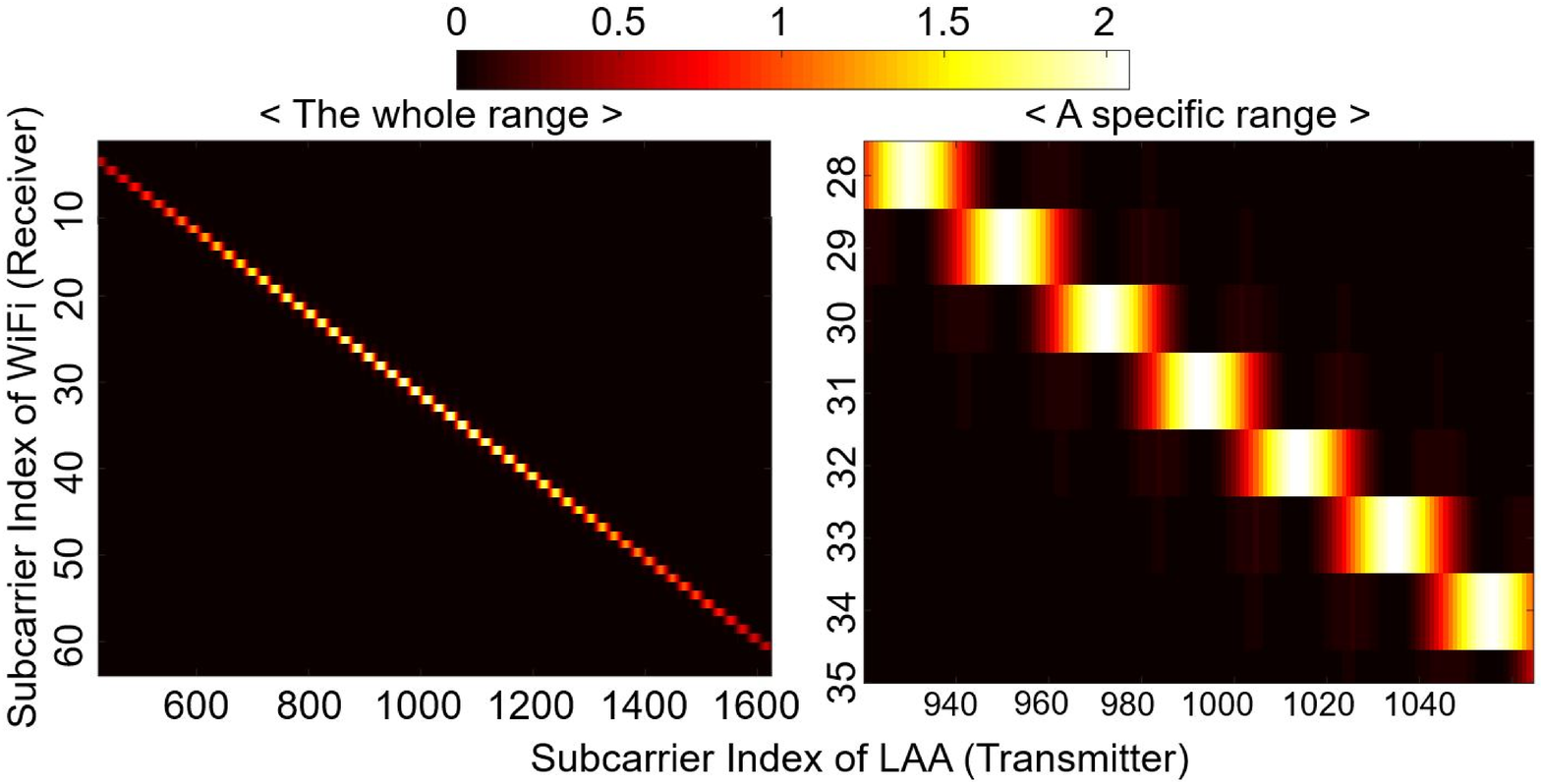}
\label{Fig:Channel_Result_LAA_to_WiFi}
}
\subfigure[Averaged squared absolute values of $\mathbf{H}_{\textrm{AP}_\alpha,\textrm{UE}}$]{
\includegraphics[width=0.97\columnwidth]{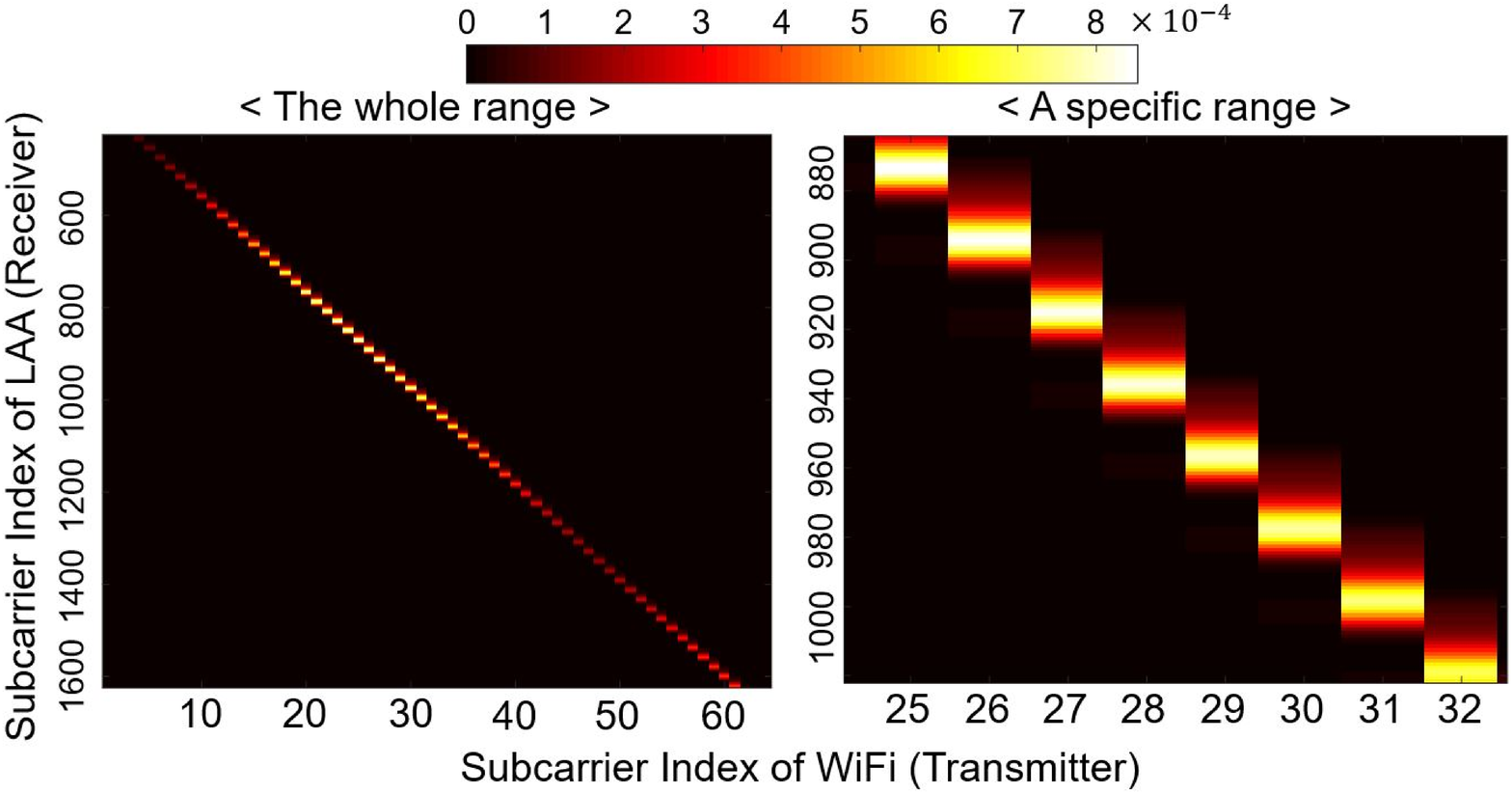}
\label{Fig:Channel_Result_WiFi_to_LAA}
}
\caption{Interference between LAA-LTE's subcarriers and Wi-Fi's subcarriers.}
\label{Fig:Channel_Result_overall}
\end{figure}
Fig.~\ref{Fig:Channel_Result_overall} presents the square of the absolute value of each element of the derived effective channel matrices, $\mathbf{H}_{\text{eNB},\text{AP}_\alpha}$ and $\mathbf{H}_{\text{AP}_\alpha,\text{UE}}$, averaged over the timing difference variables $\tau_1$ and $\tau$. Here, $\tau_1$ and $\tau$ are assumed to be uniformly distributed over $[0, 3.3~\mu s]$ and $[-T_{\text{total}}^{\text{W}},  T_{\text{total}}^{\text{L}}-T_{\text{total}}^{\text{W}}]$, respectively. 
The results show the interference from each LAA-LTE's subcarrier (or each Wi-Fi's subcarrier) to all the Wi-Fi's subcarriers (or all the LAA-LTE's subcarriers).
We consider $N_\text{tap}=16$, where the long-term gain of each channel tap exponentially decreases, and where
OFDM parameters of LAA-LTE and Wi-Fi are set in Table~\ref{Table:OFDM_Paramters}.

The left figures of Fig.~\ref{Fig:Channel_Result_LAA_to_WiFi} and Fig.~\ref{Fig:Channel_Result_WiFi_to_LAA}  show the results for all the LAA-LTE's and Wi-Fi's subcarriers, while the right ones present the enlarged results for specific ranges. 
Note that the guard subcarriers part \cite{Guard_subcarrier} of LAA-LTE is omitted from the figures.

In Fig.~\ref{Fig:Channel_Result_LAA_to_WiFi},
the whole range result presents that an LAA-LTE's subcarrier affects some Wi-Fi's subcarriers around itself.
Specifically, from the right figure, an LAA-LTE's subcarrier has the effect on one or two Wi-Fi's subcarriers.
The LAA-LTE's subcarriers close to the center frequency of a Wi-Fi's subcarrier affect mostly a single Wi-Fi's subcarrier.
As an LAA-LTE's subcarrier moves away from the center frequency of a Wi-Fi's subcarrier, however, the LAA-LTE's subcarrier starts to interfere with two Wi-Fi's subcarriers around itself.



In Fig.~\ref{Fig:Channel_Result_WiFi_to_LAA}, a similar tendency as in Fig. \ref{Fig:Channel_Result_LAA_to_WiFi} is observed. In addition, the right figure of Fig. \ref{Fig:Channel_Result_WiFi_to_LAA} presents that a Wi-Fi's subcarrier interferes with several LAA-LTE's subcarriers, which is reasonable since the subcarrier spacing of Wi-Fi is larger than that of LAA-LTE.
Specifically, a Wi-Fi's subcarrier strongly affects around 21 LAA-LTE's subcarriers.

From the simulation results, it has been confirmed that a single Wi-Fi's subcarrier affects multiple LAA-LTE's subcarriers, which should be taken into account in the  consideration of the interference between LAA-LTE and Wi-Fi. It has been also  confirmed that the accurate interference level between LAA-LTE and Wi-Fi cannot be trivially derived by only considering the difference of the subcarrier spacing of LAA-LTE and Wi-Fi, but needs a careful consideration of the symbol synchronization issue as done in Section \ref{Sec:Channel Model from an AP to an UE} and  \ref{Sec:Channel Model from an eNB to a STA}. 



\vspace{-0.05in}
\bibliographystyle{IEEEtran}
\bibliography{references}

\end{document}